\documentclass[aps,prl,twocolumn,superscriptaddress,showpacs,preprintnumbers]{revtex4}
\usepackage{graphicx} 
\usepackage{dcolumn}  
\usepackage{subfigure}
\usepackage[usenames]{color}

\graphicspath{{ps}}

\renewcommand{\overline}{\bar}
\def\be{\begin{equation}}
\def\ee{\end{equation}}

\newcommand{\lw}[1]{\smash{\lower1.7ex\hbox{#1}}}
\newcommand{\lww}[1]{\smash{\lower6.7ex\hbox{#1}}}
\newcommand{\bbbar}{\ensuremath{B\bar{B}} }

\newcommand{\btaunu}{\ensuremath{B^{-}\rightarrow\tau^{-}\bar{\nu}_\tau}}
\newcommand{\bdstlnu}{\ensuremath{B^{-}\rightarrow D^{*0}\ell^-\bar{\nu}_\ell}}
\newcommand{\btag}{\ensuremath{B_{\rm tag}}}
\newcommand{\bsig}{\ensuremath{B_{\rm sig}}}
\newcommand{\mbc}{\ensuremath{M_{\rm bc}}}
\newcommand{\de}{\ensuremath{\Delta E}}
\newcommand{\eecl}{\ensuremath{E_{\text{ECL}}}}
\newcommand{\mm}{\ensuremath{M_{\text{miss}}^2}}
\newcommand{\nb}{\ensuremath{{\cal N}_{\text{tag}}}}
\newcommand{\nsigvalue}{62}
\newcommand{\nsigstaterr}{^{+23}_{-22}}
\newcommand{\nsigsysterr}{\pm 6}
\newcommand{\brvalue}{0.72}
\newcommand{\brstaterr}{^{+0.27}_{-0.25}}
\newcommand{\brsysterr}{\pm0.11}
\newcommand{\kl}{\ensuremath{K_L^0}}
\newcommand{\ks}{\ensuremath{K_S^0}}

\newcommand{\was}{W\c as}
\begin{document}

\vspace*{-3\baselineskip}
\resizebox{!}{3cm}{\includegraphics{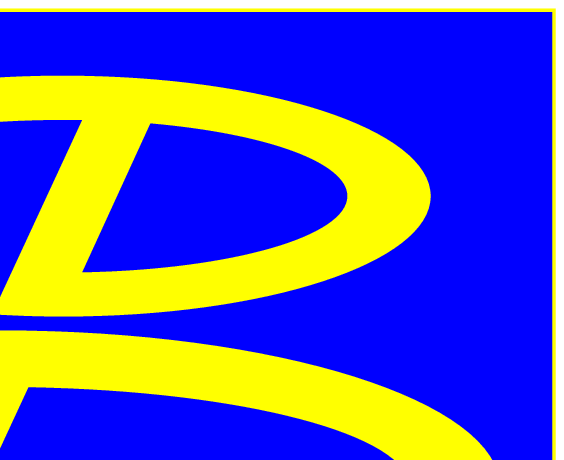}}

\preprint{\vbox{ \hbox{   }
    \hbox{Belle Preprint 2012-26}
    \hbox{KEK Preprint 2012-28}
}}

\title{ 
  \quad\\[0.5cm]
    { \boldmath Evidence for $B^- \to \tau^- \overline{\nu}_{\tau}$ with 
a Hadronic Tagging Method\\Using the Full Data Sample of Belle\\}
}

\noaffiliation

\affiliation{University of the Basque Country UPV/EHU, 48080 Bilbao}
\affiliation{University of Bonn, 53115 Bonn}
\affiliation{Budker Institute of Nuclear Physics SB RAS and Novosibirsk State University, Novosibirsk 630090}
\affiliation{Faculty of Mathematics and Physics, Charles University, 121 16 Prague}
\affiliation{University of Cincinnati, Cincinnati, Ohio 45221}
\affiliation{Justus-Liebig-Universit\"at Gie\ss{}en, 35392 Gie\ss{}en}
\affiliation{Gifu University, Gifu 501-1193}
\affiliation{II. Physikalisches Institut, Georg-August-Universit\"at G\"ottingen, 37073 G\"ottingen}
\affiliation{Hanyang University, Seoul 133-791}
\affiliation{University of Hawaii, Honolulu, Hawaii 96822}
\affiliation{High Energy Accelerator Research Organization (KEK), Tsukuba 305-0801}
\affiliation{Ikerbasque, 48011 Bilbao}
\affiliation{Indian Institute of Technology Guwahati, Assam 781039}
\affiliation{Indian Institute of Technology Madras, Chennai 600036}
\affiliation{Institute of High Energy Physics, Chinese Academy of Sciences, Beijing 100049}
\affiliation{Institute of High Energy Physics, Vienna 1050}
\affiliation{Institute of High Energy Physics, Protvino 142281}
\affiliation{Institute for Theoretical and Experimental Physics, Moscow 117218}
\affiliation{J. Stefan Institute, 1000 Ljubljana}
\affiliation{Kanagawa University, Yokohama 221-8686}
\affiliation{Institut f\"ur Experimentelle Kernphysik, Karlsruher Institut f\"ur Technologie, 76131 Karlsruhe}
\affiliation{Kavli Institute for the Physics and Mathematics of the Universe, University of Tokyo, Kashiwa 277-8583}
\affiliation{Korea Institute of Science and Technology Information, Daejeon 305-806}
\affiliation{Korea University, Seoul 136-713}
\affiliation{Kyungpook National University, Daegu 702-701}
\affiliation{\'Ecole Polytechnique F\'ed\'erale de Lausanne (EPFL), Lausanne 1015}
\affiliation{Luther College, Decorah, Iowa 52101}
\affiliation{University of Maribor, 2000 Maribor}
\affiliation{Max-Planck-Institut f\"ur Physik, 80805 M\"unchen}
\affiliation{School of Physics, University of Melbourne, Victoria 3010}
\affiliation{Moscow Physical Engineering Institute, Moscow 115409}
\affiliation{Graduate School of Science, Nagoya University, Nagoya 464-8602}
\affiliation{Kobayashi-Maskawa Institute, Nagoya University, Nagoya 464-8602}
\affiliation{Nara Women's University, Nara 630-8506}
\affiliation{National Central University, Chung-li 32054}
\affiliation{National United University, Miao Li 36003}
\affiliation{Department of Physics, National Taiwan University, Taipei 10617}
\affiliation{H. Niewodniczanski Institute of Nuclear Physics, Krakow 31-342}
\affiliation{Nippon Dental University, Niigata 951-8580}
\affiliation{Niigata University, Niigata 950-2181}
\affiliation{Osaka City University, Osaka 558-8585}
\affiliation{Pacific Northwest National Laboratory, Richland, Washington 99352}
\affiliation{Research Center for Electron Photon Science, Tohoku University, Sendai 980-8578}
\affiliation{University of Science and Technology of China, Hefei 230026}
\affiliation{Seoul National University, Seoul 151-742}
\affiliation{Sungkyunkwan University, Suwon 440-746}
\affiliation{School of Physics, University of Sydney, NSW 2006}
\affiliation{Tata Institute of Fundamental Research, Mumbai 400005}
\affiliation{Excellence Cluster Universe, Technische Universit\"at M\"unchen, 85748 Garching}
\affiliation{Tohoku Gakuin University, Tagajo 985-8537}
\affiliation{Tohoku University, Sendai 980-8578}
\affiliation{Department of Physics, University of Tokyo, Tokyo 113-0033}
\affiliation{Tokyo Institute of Technology, Tokyo 152-8550}
\affiliation{Tokyo Metropolitan University, Tokyo 192-0397}
\affiliation{Tokyo University of Agriculture and Technology, Tokyo 184-8588}
\affiliation{CNP, Virginia Polytechnic Institute and State University, Blacksburg, Virginia 24061}
\affiliation{Wayne State University, Detroit, Michigan 48202}
\affiliation{Yamagata University, Yamagata 990-8560}
\affiliation{Yonsei University, Seoul 120-749}
  \author{K.~Hara}\affiliation{High Energy Accelerator Research Organization (KEK), Tsukuba 305-0801} 
  \author{Y.~Horii}\affiliation{Kobayashi-Maskawa Institute, Nagoya University, Nagoya 464-8602} 
  \author{T.~Iijima}\affiliation{Kobayashi-Maskawa Institute, Nagoya University, Nagoya 464-8602}\affiliation{Graduate School of Science, Nagoya University, Nagoya 464-8602} 
  \author{I.~Adachi}\affiliation{High Energy Accelerator Research Organization (KEK), Tsukuba 305-0801} 
  \author{H.~Aihara}\affiliation{Department of Physics, University of Tokyo, Tokyo 113-0033} 
  \author{D.~M.~Asner}\affiliation{Pacific Northwest National Laboratory, Richland, Washington 99352} 
  \author{T.~Aushev}\affiliation{Institute for Theoretical and Experimental Physics, Moscow 117218} 
  \author{T.~Aziz}\affiliation{Tata Institute of Fundamental Research, Mumbai 400005} 
  \author{A.~M.~Bakich}\affiliation{School of Physics, University of Sydney, NSW 2006} 
  \author{M.~Barrett}\affiliation{University of Hawaii, Honolulu, Hawaii 96822} 
% ...
\collaboration{Belle Collaboration}

\begin{abstract}

We measure the branching fraction of $B^{-}\rightarrow \tau^{-}\bar{\nu}_\tau$
using the full $\Upsilon(4S)$ data sample containing 
$772\times 10^6$ \bbbar pairs collected with the Belle detector at the KEKB 
asymmetric-energy $e^{+}e^{-}$ collider.
Events with $\bbbar$ pairs are tagged by reconstructing one of the $B$ mesons
decaying into hadronic final states, and $B^-\to \tau^-\overline{\nu}_{\tau}$ 
candidates are detected in the recoil.
We find evidence for $B^{-}\rightarrow\tau^{-}\bar{\nu}_\tau$
with a significance of 3.0 standard deviations including systematic errors
and measure a branching fraction
${\cal B}(B^{-}\rightarrow\tau^{-}\overline{\nu}_{\tau}) =
[0.72^{+0.27}_{-0.25}(\text{stat})\pm 0.11(\text{syst})] \times 10^{-4}$.

\end{abstract}

\pacs{13.20.He, 14.40.Nd}

\maketitle


{\renewcommand{\thefootnote}{\fnsymbol{footnote}}}
\setcounter{footnote}{0}

The purely leptonic decay $B^{-}\rightarrow \tau^{-}\overline{\nu}_\tau$~\cite{conjugate}
is of high interest since it provides a unique opportunity to test the Standard Model (SM)
and search for new physics beyond the SM.
A recent estimate of the branching fraction based on
a global fit to the Cabibbo-Kobayashi-Maskawa (CKM) matrix elements~\cite{CKM}
is $(0.73^{+0.12}_{-0.07})\times 10^{-4}$~\cite{CKMfitter}.
In the absence of new physics, a measurement of
$B^{-}\rightarrow\tau^{-}\overline{\nu}_\tau$ provides a direct 
experimental determination of the product of the $B$ meson decay constant 
and the magnitude of the CKM matrix element $f_B|V_{ub}|$.
Physics beyond the SM, however, could significantly suppress
or enhance ${\cal B}(B^{-}\rightarrow\tau^{-}\overline{\nu}_\tau)$ 
via exchange of a new charged particle such as a charged Higgs boson 
from supersymmetry or from two-Higgs doublet models~\cite{Hou:1992sy,Baek:1999ch}.

Experimentally, it is challenging to identify the 
$B^{-}\rightarrow\tau^{-}\bar{\nu}_\tau$ decay because it involves more
than one neutrino in the final state and therefore cannot be kinematically constrained.
At $e^+e^-$ $B$ factories, one can reconstruct one of the $B$ mesons in the 
$e^+e^- \to \Upsilon(4S) \to B\overline{B}$ reaction, referred to hereafter 
as the tag side ($B_{\rm tag}$), either in hadronic decays or in semileptonic decays.
One then compares properties of the remaining particle(s), referred to as 
the signal side ($B_{\rm sig}$), to those expected for signal and background.
The method allows us to suppress strongly the combinatorial background 
from both $B\overline{B}$
and continuum $e^+e^-\rightarrow q\bar{q}$ ($q = u$, $d$, $s$, $c$) processes. 

The first evidence of $B^{-}\rightarrow\tau^{-}\bar{\nu}_\tau$
was reported by the Belle collaboration
with a significance of 3.5 standard deviations ($\sigma$)
including  systematic uncertainty
and a measured branching fraction of 
$[1.79^{+0.56}_{-0.49}(\mbox{stat})^{+0.46}_{-0.51}(\mbox{syst})] \times 10^{-4}$~\cite{Belle_Had_2006}.
This measurement used hadronic tags and a data sample
corresponding to $449 \times 10^6$ \bbbar events.
This was followed by measurements by Belle using the semileptonic 
tagging method~\cite{Belle_Semi_2010}, and also by the BaBar collaboration 
using both hadronic~\cite{BaBar_Had_2010} and semileptonic~\cite{BaBar_Semi_2010} tagging methods.
The four results are consistent.
An average branching fraction is
found to be $(1.67 \pm 0.30) \times 10^{-4}$~\cite{HFAG},
which is nearly $3\sigma$ higher than the estimate based on a global fit.
Therefore it is important to improve the precision of the measurement.

In this paper, we present a new measurement of
$B^{-}\rightarrow\tau^{-}\bar{\nu}_\tau$ using a hadronic tagging 
method and the full data sample of the Belle experiment.
The analysis described here has a number of significant improvements,
including an increased data sample (a factor of 1.7),
significantly improved hadronic tagging efficiency (a factor of 2.2),
and improved signal efficiency due to less restrictive selection requirements (a factor of 1.8).
The combined effect of these improvements and the accompanying change
in the signal to background ratio due to the looser selection criteria
results in a reduction of the expected error by a factor of two.
The new analysis has also improved systematic uncertainties.

We use a $711~\textrm{fb}^{-1}$ data sample containing $772\times 10^{6}$ 
$\bbbar$ pairs collected with the Belle detector~\cite{Belle}
at the KEKB $e^{+}e^{-}$ collider operating at the $\Upsilon(4S)$ resonance~\cite{KEKB}.
About 80\% of the data sample has been reprocessed
using improved track finding and photon reconstruction.
We use a dedicated Monte Carlo (MC) simulation based on GEANT~\cite{GEANT} to determine
the signal selection efficiency and study the background.
In order to reproduce the effect of beam background, data taken with
random triggers for each run period are overlaid on simulated events.
The $\btaunu$ signal MC events are generated by the EVTGEN package~\cite{EvtGen},
with the radiative effects based on the PHOTOS code~\cite{PHOTOS}.
To model the backgrounds from continuum processes, $b\rightarrow c$ processes,
semileptonic $b\rightarrow u$ processes, and other rare $b\rightarrow u,d,s$ processes,
we use large MC samples corresponding to 6, 10, 20, and 50 times
the integrated luminosity of the data sample, respectively.

The $\btag$ candidates are reconstructed in
615 exclusive charged $B$ meson decay channels using an improved
full-reconstruction algorithm~\cite{FullRecon}.
An output full-reconstruction-quality variable $\nb$ ranges from zero
for combinatorial background and continuum events 
to unity if an unambiguous $\btag$
is obtained from the hierarchical neural network.
We also use the energy difference $\de = E_{\btag} - E_{\rm CM}/2$ and
the beam-energy-constrained mass $\mbc = \sqrt{(E_{\rm CM}/2)^2/c^4 - |\vec{p}_{\btag}|^2/c^2}$,
where $E_{\rm CM}$ is the $e^+e^-$ center-of-mass (CM) energy,
and $E_{\btag}$ and $\vec{p}_{\btag}$ are the energy and the momentum,
respectively, of the $\btag$ candidate defined in the CM frame.
Charged $\btag$ candidates with 
$\nb>0.03$, $-0.08~\mbox{GeV}<\Delta E< 0.06~\mbox{GeV}$,
and $5.27~\mbox{GeV}/c^{2} < M_{\rm bc} < 5.29~\mbox{GeV}/c^{2}$ are selected.
The tag efficiency (0.24\%) and the purity (65\%) are improved
by factors of 1.7 and 1.2, respectively, compared to Ref.~\cite{Belle_Had_2006}.
The number of $\btag$'s obtained for the full data set is $1.8\times 10^6$.
In the case of $\btaunu$ signal,
in which the $\bbbar$ event has lower than average particle multiplicity,
the tag efficiency is 0.31\%.
This tag efficiency is 2.2 times higher
than that in the previous analysis~\cite{Belle_Had_2006}.

In events where $\btag$ candidates are reconstructed,
we search for $\btaunu$ decays.
The $\tau^-$ lepton is identified in the $e^{-}\bar{\nu}_{e}\nu_\tau$,
$\mu^{-}\bar{\nu}_{\mu}\nu_\tau$, $\pi^{-}\nu_{\tau}$, and $\pi^{-}\pi^{0}\nu_{\tau}$ decay channels.
Candidate events are required to have one track
with charge opposite that of the $\btag$ candidate.
The charged tracks are required to satisfy $dz < 3~{\rm cm}$ and $dr < 0.5~{\rm cm}$,
where $dz$ and $dr$ are unsigned impact parameters relative to the interaction point
along and perpendicular to the beam axis, respectively.
Charged tracks are classified as electron, muon, and pion candidates
after rejecting kaon and proton candidates~\cite{Belle}.
Candidate $\tau^- \to \pi^{-}\pi^{0}\nu_{\tau}$ events are required
to have one $\pi^0$ candidate reconstructed from $\pi^0\to\gamma\gamma$
in which neither daughter photon was used in the $\btag$ reconstruction.
The invariant mass of the $\pi^-\pi^0$ state is required to be within 0.15~GeV
of the nominal $\rho^-$ mass~\cite{PDG}.
Multiple neutrinos in the final state are distinguished
using the missing mass squared variable
$\mm = ( E_{\rm CM} - E_{\btag} - E_{\bsig} )^2/c^4 - |\vec{p}_{\btag} + \vec{p}_{\bsig}|^2/c^2$,
where $E_{\bsig}$ and $\vec{p}_{\bsig}$
are the energy and the momentum, respectively, of the $\bsig$ candidate in the CM frame.
To avoid potential backgrounds from $e^-\bar{\nu}_e$, $\mu^-\bar{\nu}_\mu$, $\pi^-K_L^0$, and $\rho^-K_L^0$,
we require $\mm > 0.7~{\rm GeV}^2/c^4$.

After removing the particles from the $\btag$ candidate
and the charged tracks and $\pi^0$'s from the $\bsig$ candidate,
there should be no other detected particles.
We require that there be no extra charged tracks
with $dz < 75~{\rm cm}$ and $dr < 15~{\rm cm}$
nor extra $\pi^0$ candidates (``$\pi^0$ veto'')
nor $\kl$ candidates (``$\kl$ veto'').
The $\kl$ veto is based on the hit patterns in the $\kl$ detection system~\cite{Belle}
that are not associated with any charged tracks.
We define the extra energy $\eecl$~\cite{Belle_Had_2006},
which is the sum of the energies of neutral clusters detected in the electromagnetic calorimeter
that are not associated with either the $\btag$
or the $\pi^0$ candidate from the $\tau^-\rightarrow \pi^{-}\pi^{0}\nu_{\tau}$ decay.
The signal has either zero or a small value of $\eecl$,
while background events tend to have larger values
due to the contributions from additional neutral clusters.
The selection criteria for $\btag$ and extra charged tracks
are optimized to maximize the sensitivity in a signal enhanced region $\eecl < 0.2~{\rm GeV}$.
We retain candidate events in the range $\eecl < 1.2~{\rm GeV}$,
where the correlation between $\eecl$ and $\mm$ is small for each background component.

The signal detection efficiency is estimated based on MC samples
after applying a correction for the $\btag$ reconstruction efficiency.
The correction factor is obtained by fitting the $\mbc$ distribution
for an $\eecl$ sideband sample defined by $0.4~{\rm GeV}<\eecl<1.2~{\rm GeV}$,
for which the kinematics is expected to be similar to the signal.
The resulting efficiencies are summarized in Table~\ref{tab:result}.
The validity of the efficiency estimation is checked by using
a semileptonic decay sample in which $\bsig$ is reconstructed
in the decay chain $\bdstlnu$ ($\ell = e$ or $\mu$) followed
by $D^{*0}\to D^0\pi^0$ and $D^0 \to K^- \pi^+$.

The signal yield is extracted from a two-dimensional
extended maximum likelihood fit to $\eecl$ and $\mm$.
The likelihood is
\begin{equation}
{\cal L} = \frac{e^{-\sum_j n_j}}{N!}
\prod_{i=1}^{N}\sum_j n_j f_j(E_{i},M^2_{i}),
\end{equation}
where $j$ is an index for the signal and background contributions, 
$n_{j}$ and $f_{j}$ are the yield and the probability density
function (PDF), respectively, of the $j^{\textrm{th}}$ contribution,
$E_{i}$ and $M^2_{i}$ are the $\eecl$ and $\mm$ values in the $i^{\textrm{th}}$ event, respectively,
and $N$ is the total number of events in the data.
The signal component in $\tau^- \to \pi^-\nu_{\tau}$ candidate events
includes large cross-feed contributions from $\tau^- \to \ell^{-}\bar{\nu}_{\ell}\nu_\tau$
and $\tau^- \to \pi^- \pi^0\nu_{\tau}$ decays. 
The dominant background contribution is from $b\rightarrow c$ decays. 
The small backgrounds from charmless $B$ decays
and continuum processes are also included in the fit.
In the final sample,
the fractions of the backgrounds from $b\rightarrow c$ decays,
charmless $B$ decays, and continuum processes are estimated from MC
to be 89.8\%, 9.7\%, and 0.5\% for leptonic $\tau^-$ decays and 
75.1\%, 6.5\%, and 18.4\% for hadronic $\tau^-$ decays.
The PDFs are constructed by taking products of one-dimensional histograms
in $\eecl$ and $\mm$ obtained from MC for all contributions
except for cross-feed from $\tau^- \to \pi^- \pi^0\nu_{\tau}$ decays
in $\tau^- \to \pi^-\nu_{\tau}$ candidate events;
for this component a two-dimensional histogram PDF is used
to take into account the correlation originating from the misreconstructed $\pi^0$.

The $B$ decays in which only one charged particle is detected
can make a peak near zero $\eecl$ and mimic the signal.
These are predominantly
$B^- \to D^{(*)0}\ell^-\overline{\nu}_\ell$ and $\bar{B}^0 \to D^{(*)+}\ell^-\overline{\nu}_\ell$ decays,
where the $D$ decays semileptonically or to a final state with one or more $\kl$'s.
Charmless $B$ decays such as $B^- \to \pi^0 \ell^-\bar{\nu}_{\ell}$, $K^-\nu\overline{\nu}$, 
$\kl\pi^-$, $K^{*-}\gamma$, and $\mu^- \overline{\nu}_{\mu}\gamma$ can also contribute.
The fraction in the signal enhanced region $\eecl < 0.2~{\rm GeV}$
of these peaking decay modes over the total background is 32\% according to the MC simulation.

\begin{figure}[htbp]
 \begin{center}
  \leavevmode
  \subfigure
  {\includegraphics[width=0.22\textwidth]{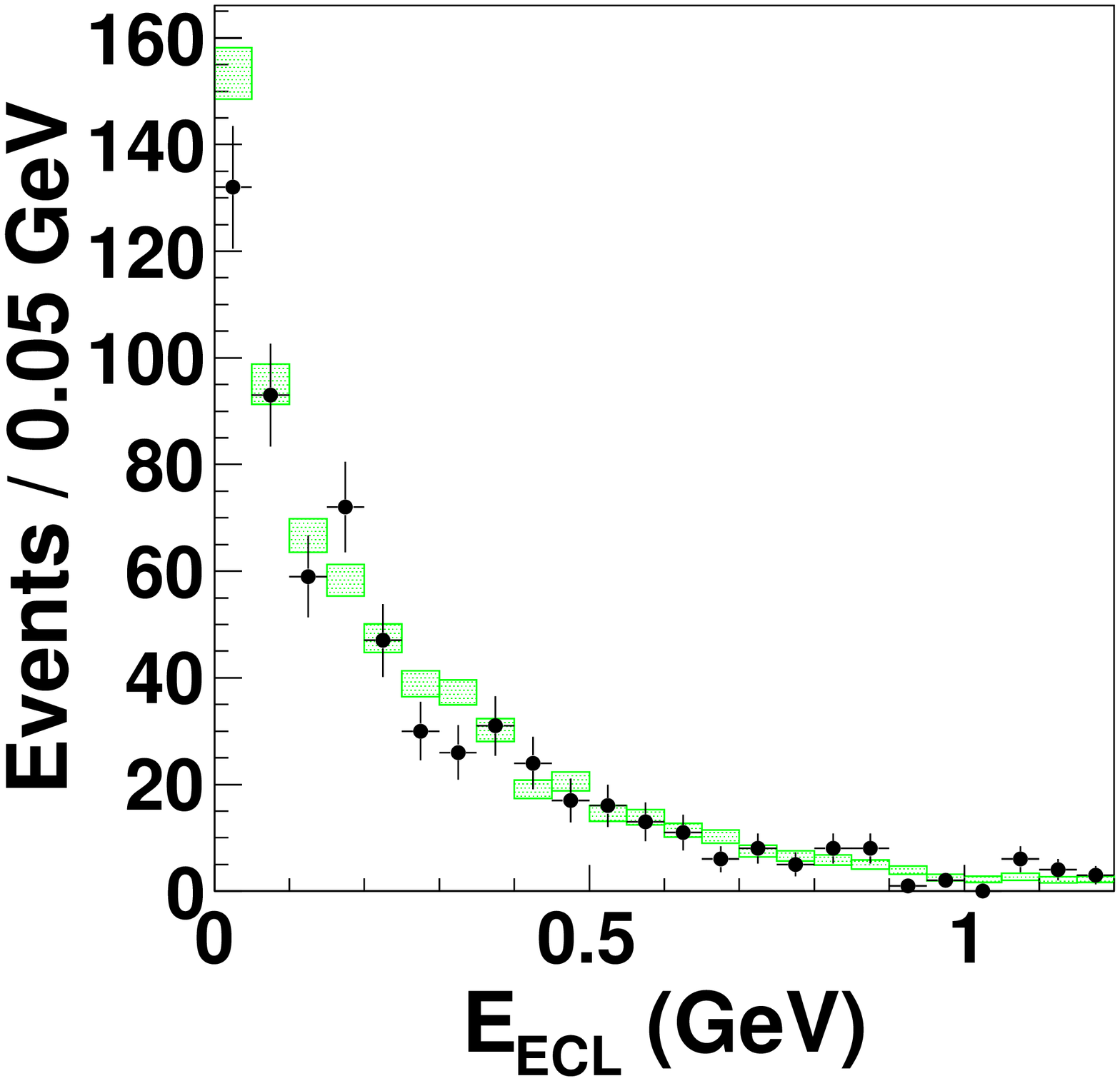}}
  \hspace{2mm}
  \subfigure
  {\includegraphics[width=0.22\textwidth]{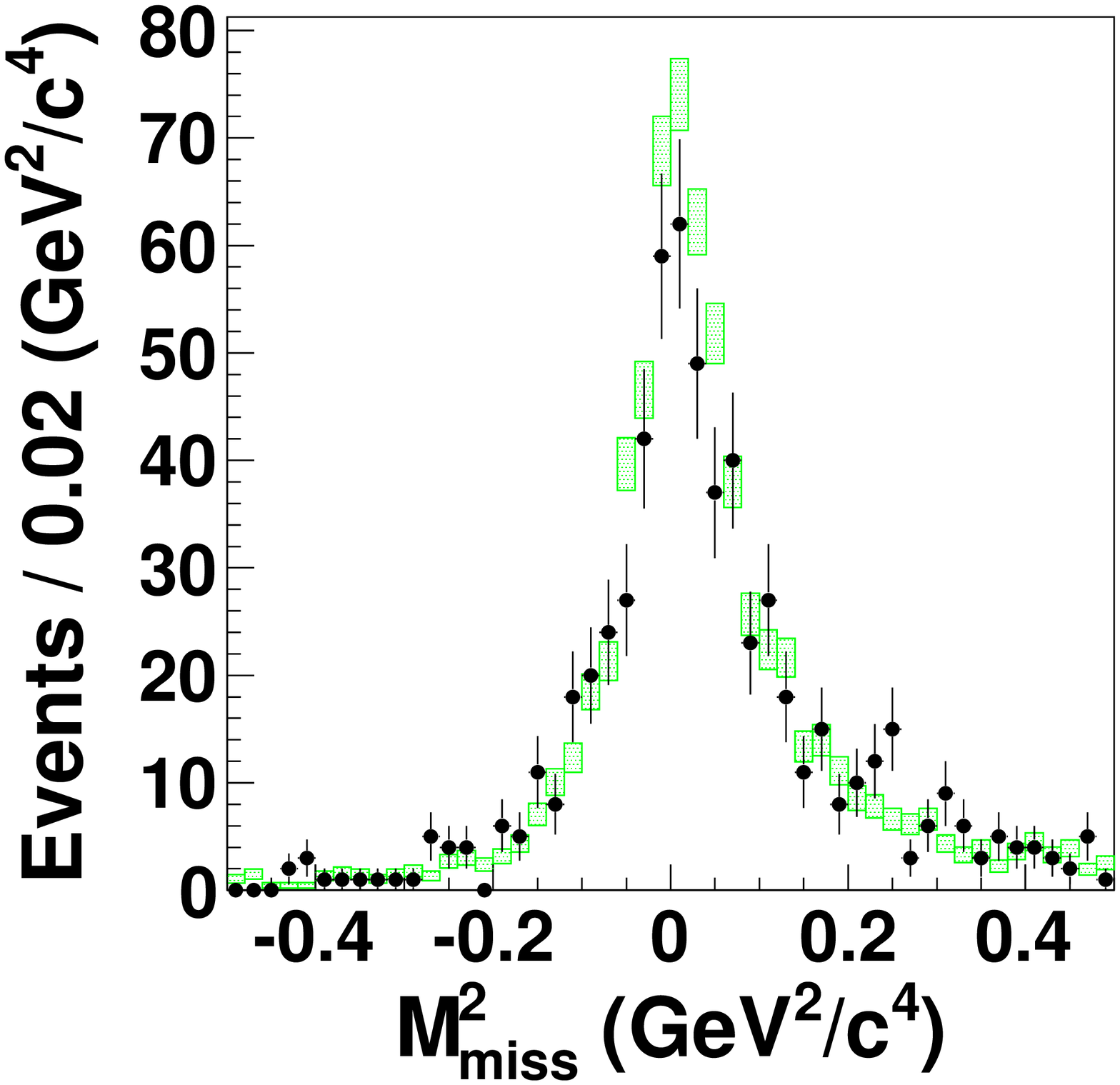}}
  \vspace{-2mm}
  \caption{
  Distributions of $\eecl$ (left) and $\mm$ (right) for $\bdstlnu$.
  The dots with error bars show the data.
  The rectangles show the normalized MC simulation,
  where the MC size is five times larger than the data.}
  \label{fig:dstlnu}
 \end{center}
\end{figure}

The simulated $\eecl$ and $\mm$ distributions in MC are
validated using various control samples.
A non-zero $\eecl$ value for the $\btaunu$ signal component is
due to beam background and split-off showers originating from $\btag$ and $\bsig$ decay products.
The average contributions from these sources are 0.04 GeV, 0.12 GeV,
and 0.08 GeV, respectively, per event in the signal MC sample.
The simulated $\eecl$ distribution is checked
with the $\bdstlnu$ sample, which has a final state
similar to the $\btaunu$ signal if the $D^{*0}$ decay products are removed.
We also check the difference between the detector resolution
in data and MC for $\mm$ with the $\bdstlnu$ sample.
We confirm that the $\eecl$ distributions and $\mm$ resolutions of data and MC are consistent 
for the $\bdstlnu$ sample as shown in Figure~\ref{fig:dstlnu}.
The background $\eecl$ and $\mm$ descriptions by MC are checked using sidebands
in $\mbc$ and $\eecl$, events with the $\btag$ reconstructed in a $B^0$ mode,
and events with the same $\bsig$ charge as the $\btag$.
The $\kl$ detection efficiency is
calibrated using a $D^0\to\phi\ks$ data sample
by comparing the yields of $\phi\to\kl\ks$ and $\phi\to K^+K^-$ decays.
We confirm the MC expectations for the $\eecl$ and $\mm$ shapes
and verify that the normalization agrees with data 
after the calibrations of the $\btag$ and $\kl$ reconstruction efficiencies.

In the final fit, five parameters are allowed to vary: the total signal yield
and the sum of the backgrounds from $b\rightarrow c$ decays
and continuum processes for each $\tau^-$ decay mode.
The ratio of the $b\rightarrow c$ and continuum backgrounds is fixed to the value obtained from MC
after the $\btag$ efficiency correction has been applied.
The background contributions from charmless $B$ decays are fixed to the MC expectation.
We combine $\tau^-$ decay modes by constraining the ratios of the signal 
yields to the ratios of the reconstruction efficiencies obtained from MC
including the branching fractions of $\tau^-$ decays~\cite{PDG}.

\begin{figure}[htbp]
 \begin{center}
  \leavevmode
  \subfigure
  {\includegraphics[width=0.38\textwidth]{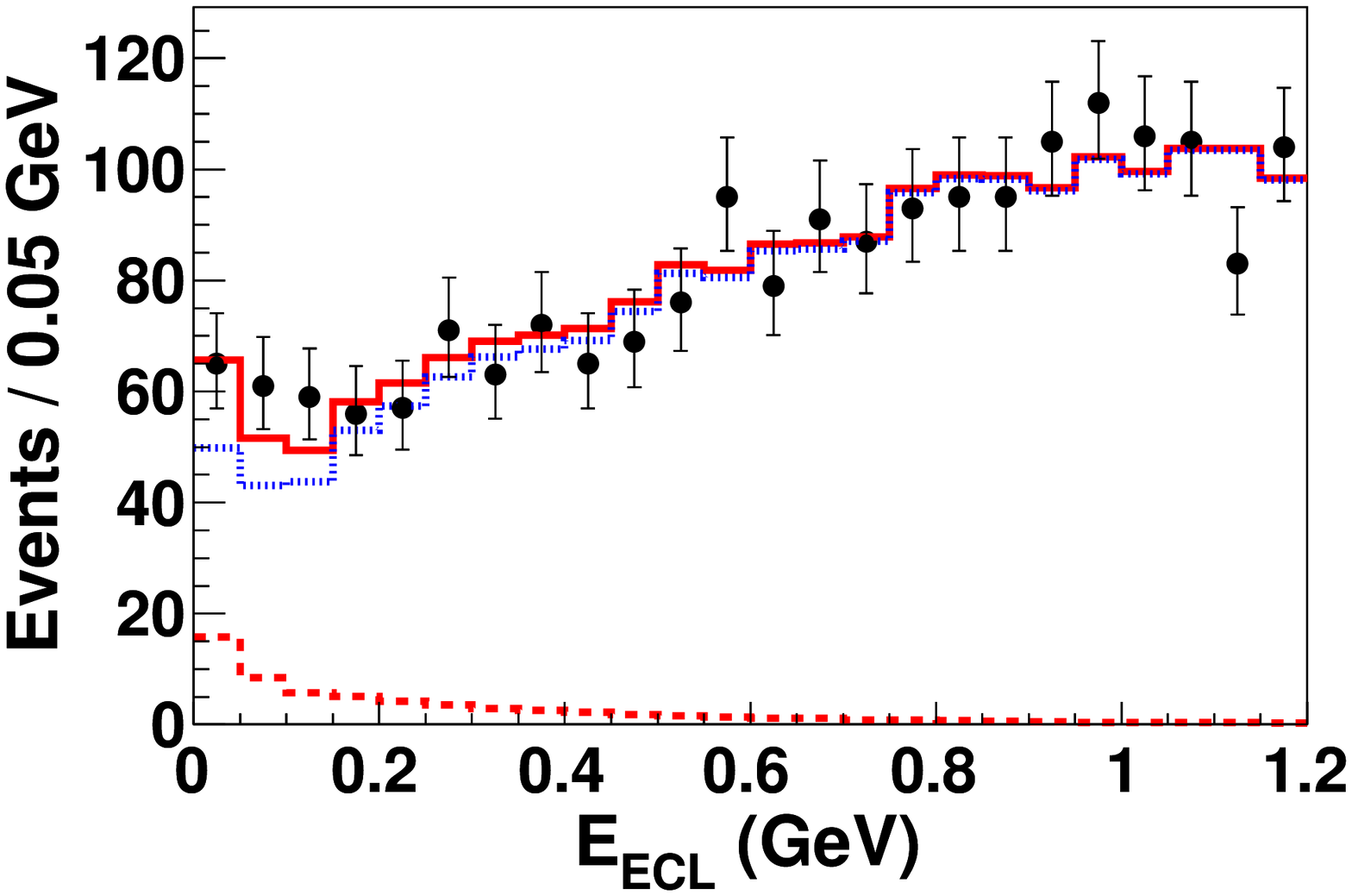}}\\
  \vspace{-3mm}
  \subfigure
  {\includegraphics[width=0.38\textwidth]{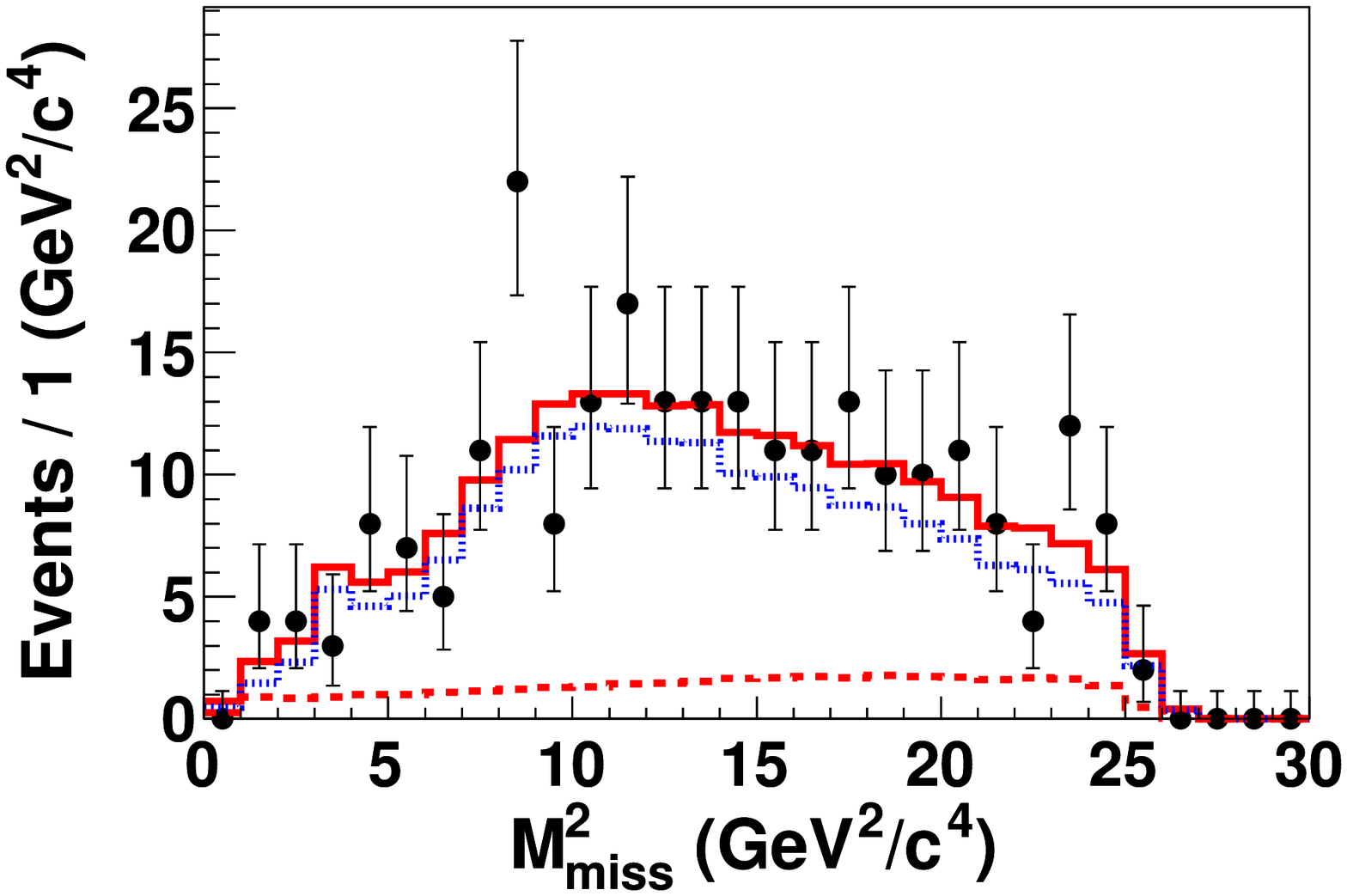}}\\
  \vspace{-3mm}
  \caption{
  Distributions of $\eecl$ (top) and $\mm$ (bottom) combined for all the $\tau^-$ decays.
  The $\mm$ distribution is shown for a signal region of $\eecl < 0.2~{\rm GeV}$.
  The solid circles with error bars are data.
  The solid histograms show the projections of the fits.
  The dashed and dotted histograms show the signal and background components, respectively.
  }
  \label{fig:result}
 \end{center}
\end{figure}

Figure~\ref{fig:result} shows the result of the fit to the $\eecl$ and $\mm$ distributions
for all the $\tau^-$ decay modes combined.
The signal yield is $\nsigvalue\nsigstaterr({\rm stat})\nsigsysterr({\rm syst})$,
where the first and second errors correspond
to statistical and systematic uncertainties, respectively.
The significance of the signal is estimated
by $\sqrt{-2\ln({\cal L}_0/{\cal L}_{\rm max})}$,
where ${\cal L}_{\rm max}$ and ${\cal L}_0$ are the maximum likelihood
and the likelihood obtained assuming zero signal yield, respectively.
The likelihoods are obtained after convolving with a Gaussian distribution
that corresponds to the systematic error.
We obtain a significance of 3.0$\sigma$
including systematic uncertainties.
The branching fraction is calculated
by ${\cal B} = N_{\rm sig} / (2 \epsilon N_{B^+B^-})$,
where $N_{\rm sig}$ is the signal yield, $\epsilon$ is the efficiency,
and $N_{B^+B^-}$ is the number of $B^+B^-$ events.
Equal production of neutral and charged $B$ meson pairs
in $\Upsilon(4S)$ decay is assumed. We obtain
\begin{equation}
{\cal B}(\btaunu) = [\brvalue\brstaterr({\rm stat})\brsysterr({\rm syst})]\times 10^{-4}.
\end{equation}
The result is summarized in Table~\ref{tab:result}.

As a check, we fit the $\eecl$ and $\mm$ distributions
while floating the yield for each of the four $\tau^-$ decay modes.
The resulting yields, as well as the efficiencies and the branching fractions,
are listed in Table~\ref{tab:result}.
We include the $e^-\bar{\nu}_e\nu_\tau$, $\mu^-\bar{\nu}_\mu\nu_\tau$,
and $\pi^- \pi^0 \nu_\tau$ cross-feeds in the $\pi^- \nu_\tau$ candidate events
in the $e^-\bar{\nu}_e\nu_\tau$, $\mu^-\bar{\nu}_\mu\nu_\tau$,
and $\pi^- \pi^0 \nu_\tau$ signal yields.
The branching fractions are in good agreement between different $\tau^-$ decays.
We also check the result after removing the $\kl$ veto,
and obtain $N_{\rm sig} = 65^{+27}_{-25}(\text{stat})$ and
${\cal B}(\btaunu) = [0.65^{+0.27}_{-0.25}(\text{stat})] \times 10^{-4}$.
These checks are consistent with the nominal result.
In addition, we perform one-dimensional fits
to $\eecl$ and $\mm$ and divide the data sample into several subsets.
All results are in good agreement with the nominal result
within the statistical errors.

\begin{table}[htb]
 \caption{
 Results of the fit for $\btaunu$ yields~($N_{\rm sig}$), detection efficiencies~($\epsilon$),
 and branching fractions~($\cal B$).
 The efficiencies include the branching fractions of the $\tau^-$ decay modes.
 The errors for $N_{\rm sig}$ and $\cal B$ are statistical only.
 }
 \label{tab:result}
 \begin{center}
  \begin{tabular}{lccc}
  \hline \hline
  Sub-mode & $N_{\rm sig}$ & $\epsilon$ ($10^{-4}$) & $\cal B$ ($10^{-4}$) \\ \hline
  $\tau^-\rightarrow e^{-}\bar{\nu}_{e}\nu_\tau$ 
  & $16^{+11}_{-9}$ & 3.0 & $0.68^{+0.49}_{-0.41}$ \\
  $\tau^-\rightarrow \mu^{-}\bar{\nu}_{\mu}\nu_\tau$ 
  & $26^{+15}_{-14}$ & 3.1 & $1.06^{+0.63}_{-0.58}$ \\
  $\tau^-\rightarrow \pi^{-}\nu_\tau$
  & $8^{+10}_{-8}$ & 1.8 & $0.57^{+0.70}_{-0.59}$ \\
  $\tau^-\rightarrow \pi^-\pi^0\nu_\tau$
  & $14^{+19}_{-16}$ & 3.4 & $0.52^{+0.72}_{-0.62}$ \\
  Combined
  & $62^{+23}_{-22}$ & 11.2 & $0.72^{+0.27}_{-0.25}$ \\
  \hline \hline
  \end{tabular}
 \end{center}
\end{table}

Systematic errors for the measured branching fraction are associated with 
the uncertainties in the signal yield, the efficiencies, and the number of $B^+B^-$ pairs.
The systematic error from MC statistics of the PDF histograms 
is evaluated by varying the content of each bin by its statistical uncertainty.
To estimate the systematic error due to the possible signal $\eecl$ 
shape difference between MC and data,
the ratio of data to MC for the $\eecl$ histograms of the $\bdstlnu$
sample is fitted with a first-order polynomial and 
the signal $\eecl$ PDF is modified within the fitted errors.
The uncertainties for the branching fractions of the $B$
decays that peak near zero $\eecl$ are estimated by changing 
the branching fractions in MC by their experimental errors~\cite{PDG} 
if available, or by $\pm50$\% otherwise.
The sizes of these backgrounds also depend on the fractions
of the events with correctly reconstructed $\btag$,
and related systematic uncertainties are obtained by
using the statistical errors for the fractions in the MC simulation.
To estimate the uncertainty associated with the $\btag$ efficiency 
for the signal, ${\cal B}(\bdstlnu)$ obtained from the $\bdstlnu$ sample 
is compared to the world average value~\cite{PDG}. The results are consistent and 
the uncertainty of the measurement is assigned as the systematic error.
The systematic errors in the signal-side efficiencies arise from the uncertainty
in tracking efficiency, particle identification efficiency, $\pi^0$ reconstruction efficiency,
branching fractions of $\tau^-$ decays, and MC statistics.
The systematic uncertainty related to the $\kl$ veto efficiency 
is estimated from the statistical uncertainties of the $D^0 \to \phi \ks$ control sample
and the fraction of events with $\kl$ candidates in the $\bdstlnu$ sample.
The total systematic error is calculated
by summing the above uncertainties in quadrature.
The estimated systematic errors are summarized in Table~\ref{tab:syst}.

\begin{table}[htb]
 \caption{
   Summary of the systematic errors for the $\btaunu$ branching fraction measurement.
 }
 \label{tab:syst}
 \begin{center}
  \begin{tabular}{lc}
  \hline \hline
  Source & $\cal B$ syst. error (\%) \\ \hline 
  Signal PDF & 4.2 \\
  Background PDF & 8.8 \\
  Peaking background & 3.8 \\
  $\btag$ efficiency & 7.1 \\
  Particle identification & 1.0 \\
  $\pi^0$ efficiency & 0.5 \\
  Tracking efficiency & 0.3 \\
  $\tau$ branching fraction & 0.6 \\
  MC efficiency statistics & 0.4 \\ 
  $\kl$ efficiency & 7.3 \\
  $N_{B^+B^-}$ & 1.3 \\ \hline
  Total & 14.7 \\
  \hline \hline
  \end{tabular}
 \end{center}
\end{table}

The branching fraction measured here is lower than the previous Belle result
with a hadronic tagging method~\cite{Belle_Had_2006}.
Using the first sample of $449 \times 10^6 \bbbar$ pairs, which corresponds to the
data set used in Ref.~\cite{Belle_Had_2006} after reprocessing,
we obtain ${\cal B}(\btaunu) = [1.08^{+0.37}_{-0.35}(\text{stat})] \times 10^{-4}$.
Note that 89\% of the events in the final sample in this analysis
is not included in the final sample in Ref.~\cite{Belle_Had_2006}
mainly due to the loosened selection, the different $\btag$ reconstruction method, and the $\kl$ veto.
Using the last $323 \times 10^6 \bbbar$ pairs, we obtain
${\cal B}(\btaunu) = [0.24^{+0.39}_{-0.34}(\text{stat})] \times 10^{-4}$,
which is statistically consistent with the result for the first 
$449 \times 10^6 \bbbar$ data set within 1.6$\sigma$.
Our results are also consistent with other publications
within the errors~\cite{Belle_Semi_2010, BaBar_Had_2010, BaBar_Semi_2010}.

In summary, we measure the branching fraction of the decay 
$B^{-}\rightarrow\tau^{-}\overline{\nu}_\tau$ with hadronic tagging
using Belle's final data sample containing $772 \times 10^6$ \bbbar pairs.
We find evidence for
$B^{-}\rightarrow\tau^{-}\overline{\nu}_\tau$
with a signal significance of $3.0\sigma$ including systematic uncertainties
and measure a branching fraction of
$[\brvalue\brstaterr({\rm stat})\brsysterr({\rm syst})]\times 10^{-4}$.
By employing a neural network-based method for hadronic tagging
and a two-dimensional fit for signal extraction, along with
a larger data sample, both statistical and systematic precisions
are significantly improved compared to the previous analysis~\cite{Belle_Had_2006}.
The result presented in this paper supersedes the previous result 
reported in Ref.~\cite{Belle_Had_2006}.
Combined with the Belle measurement
based on a semileptonic $B$ tagging method~\cite{Belle_Semi_2010}
taking into account all the correlated systematic errors,
the branching fraction is found to be 
${\cal B}(B^{-}\rightarrow\tau^{-}\overline{\nu}_{\tau}) =
(0.96 \pm 0.26) \times 10^{-4}$,
with a $4.0\sigma$ signal significance including systematic uncertainties.
This value is consistent with the SM expectation
obtained from other experimental constraints. 
Using this result and parameters found in Ref.~\cite{PDG}, we obtain
$f_B |V_{ub}|= [7.4 \pm 0.8({\rm stat}) \pm 0.5({\rm syst})] \times 10^{-4}~{\rm GeV}$.
Our result provides stringent constraints on various models of new physics
including charged Higgs bosons.

\begin{acknowledgements}


We thank the KEKB group for excellent operation of the
accelerator; the KEK cryogenics group for efficient solenoid
operations; and the KEK computer group, the NII, and 
PNNL/EMSL for valuable computing and SINET4 network support.  
We acknowledge support from MEXT, JSPS and Nagoya's TLPRC (Japan);
ARC and DIISR (Australia); NSFC (China); MSMT (Czechia);
DST (India); INFN (Italy); MEST, NRF, GSDC of KISTI, and WCU (Korea); 
MNiSW (Poland); MES and RFAAE (Russia); ARRS (Slovenia); 
SNSF (Switzerland); NSC and MOE (Taiwan); and DOE and NSF (USA).
We are grateful for the support of JSPS KAKENHI Grant Number 24740157.

\end{acknowledgements}


\begin{thebibliography}{99}
\bibitem{conjugate}
Charge-conjugate decays are implied throughout this paper unless otherwise stated.
\end{thebibliography}
\end{document}